\documentclass[a4paper,fleqn,usenatbib]{mnras}

\usepackage{newtxtext,newtxmath}

\usepackage[T1]{fontenc}
\usepackage{ae,aecompl}

\usepackage{epstopdf}
\usepackage{ulem}
\usepackage{xcolor}
\usepackage{multirow}
\usepackage{graphicx}	
\usepackage{amsmath}	
\usepackage{amssymb}	

\title{Magnetic ionization-thermal instability}

\author[A.~E.~Dudorov, C.~E.~Stepanov, , S.~O.~Fomin, S.~A.~Khaibrakhmanov]{
A.~E. Dudorov$^{1}$\thanks{E-mail: dudorov@csu.ru (AED)},
C.~E. Stepanov$^{1}$,
S.~O. Fomin$^{1}$\thanks{E-mail: fominso@csu.ru (SOF)},
and S.~A. Khaibrakhmanov$^{1,2}$\thanks{E-mail: khaibrakhmanov@csu.ru (SAKh)}
\\
$^{1}$Theoretical   Physics Department,
Chelyabinsk State University,  Br. Kashirinykh  St., 129, Chelyabinsk,
454001, Russia\\
$^{2}$ Ural Federal University, 51 Lenin str., Ekaterinburg 620000, Russia\\
}

\date{Accepted XXX. Received YYY; in original form ZZZ}

\pubyear{2019}

\begin{document}
\label{firstpage}
\pagerange{\pageref{firstpage}--\pageref{lastpage}}
\maketitle

\begin{abstract}
Linear analysis of the stability of diffuse clouds in the cold neutral medium with uniform magnetic field is performed. We consider that gas in equilibrium state is heated by cosmic rays, X-rays and electronic photoeffect on the surface of dust grains, and it is cooled by the collisional excitation of fine levels of the C~\textsc{ii}. Ionization by cosmic rays and radiative recombinations is taken into account. A dispersion equation is solved analytically in the limiting cases of small and large wave numbers, as well as numerically in the general case.
In particular cases the dispersion equation describes thermal instability of Field (1965) and ionization-coupled acoustic instability of Flannery and Press (1979). We pay our attention to magnetosonic waves arising in presence of magnetic field, in thermally stable region, $35 \leq T \leq 95$~K and density $n\la 10^3\,\mbox{cm}^{-3}$. We have shown that these modes can be unstable in the isobarically stable medium. The instability mechanism is similar to the mechanism of ionization-coupled acoustic instability.
We determine maximum growth rates and critical wavelengths of the instability of magnetosonic waves depending on gas temperature, magnetic field strength and the direction of wave vector with respect to the magnetic field lines. The minimum growth time of the unstable slow magnetosonic waves in diffuse clouds is of $4-60$~Myr, minimum and the most unstable wavelengths lie in ranges $0.05-0.5$ and $0.5-5$~pc, respectively. We discuss the application of considered instability to the formation of small-scale structures and the generation of MHD turbulence in the cold neutral medium.

\end{abstract}

\begin{keywords}
ISM: clouds -- instabilities -- magnetic fields
\end{keywords}



\section{Introduction}

The interstellar medium (ISM) in the Galaxy consists of three phases in a state of dynamical equilibrium~\citep{mckee}: the cold neutral medium (CNM) with $T<100$~K, the warm intercloud medium with $T\approx 8000$~K, and the hot coronal gas  with $T\approx 10^6$~K. There is thermal equilibrium between the cold and warm phases. Measurements of Faraday rotation and dispersion measurements of pulsars give mean magnetic field strength of $2\,\umu$G in the solar neighborhood ~\citep{ruzmaikin77, inoue81}.
According to recent data on synchrotron polarization, the Galactic magnetic field strength is up to $\sim 5\,\umu$G \citep{planck16_XLII}.
Zeeman-splitting measurements of the 21 cm absorption line show that  median value of the magnetic field is of $6\,\umu$G in diffuse clouds with density $\sim 10-100\,\mbox{cm}^{-3}$~\citep{heiles05,crutcher10,heiles12}.
These values correspond to plasma beta $\la 1$, therefore the magnetic field plays important  role in the dynamics of the ISM.

The CNM is non-homogeneous \citep[see review by][]{snow06}. The densest parts of the CNM are the complexes of molecular clouds or giant molecular clouds (GMC). GMC demonstrate filamentary structure, i.e. they have the form of filaments and sheets~\citep[see review][]{andre14}.  The CNM also contains translucent clouds with  densities $n\sim 500\,\rm{cm}^{-3}$ and temperatures $15-30$~K, and diffuse clouds with $n=10-500\,\rm{cm}^{-3}$ and temperatures $30-100$~K. Moreover, the tiny scale atomic structures (TSAS) are observed in the CNM with typical sizes $10-1000$~au, $n=10^3-10^4\,\rm{cm}^{-3}$, $T=15-100$~K~\citep{dieter76, heiles, stanimirovic, stanimirovic18}. Nature of the small-scale structures in the CNM is still under debate.

The formation of various structures in diffuse ISM is explained by the action of different instabilities and turbulence under the influence of the magnetic field~\citep[see reviews][]{dudorov91, elmegreen04, hennebelle2009, dudorov17}. Separation of the ISM into cold clouds surrounded by warm gas is caused by thermal instability~(\citealt{field1}; \citealt{pikelner67} with English translation \citealt{pikelner68}; \citealt{field2}).  The turbulence in the ISM arises due to Kelvin-Helmholtz instability that may develop in the converging flows from supernova remnants.
The structures in diffuse ISM can be a product of the interplay between compressible turbulence and heating and cooling processes in the neutral interstellar gas. Some fraction of thermally unstable gas can be produced by turbulent motions. The collision of turbulent flows can initiate condensation of warm neutral medium into cold neutral clouds with the fraction of cold gas, as well as the fraction of thermally unstable gas~\citep[see][and references therein]{banerjee09, hennebelle2009}.

Thermal instability develops usually as isobaric, isochoric and
isentropic modes.
The unstable isobaric and isochoric modes are dynamical ones. The unstable isentropic mode is
either dynamical or overstable \citep{field1}.

Various applications of thermal instability were considered in a number of papers.
\citet{defouw} have shown
that the partially ionized hydrogen that is cooled by free-bound emissions can be thermally unstable.
Frozen-in magnetic field decreases the increments of the isobaric thermal instability in the direction nonparallel to the magnetic field lines \citep{heyvaerts}.
Ohmic and ambipolar diffusion weaken
the influence
of magnetic field on the isobaric mode \citep{heyvaerts,massaglia,ghanbari,stiele}.
The friction between ions and neutrals can weaken thermal instability in two-fluid plasma
with magnetic field \citep{fukue}.

Several authors investigated the isentropic instabilities in the ISM.
\citet{oppenheimer} discussed general conditions for the isentropic instability
in molecular clouds.
\citet{nakariakov} investigated the isentropic instability in weak non-linear approximation.
The isentropic instability can develop in the atomic zone of
photodissociation regions~\citep{krasnobaev}.

\citet{yoneyama} and \citet{glassgold} investigated thermal-chemical instability.
In the case of ionization/recombination
processes this instability was called
as a thermo-reactive instability \citep{yoneyama,corbelli}.

\citet{flannery} investigated
ionization-coupled acoustic instability in the cold diffuse ISM. They have shown that thermally stable gas can be acoustically unstable, and discussed the applications of this instability to star formation.
This instability has the following physical mechanism.
Let an element of gas is compressed by a sound wave on the intermediate
time-scale between the cooling and recombination time-scales in the thermally stable medium. If the
cooling time-scale is shorter than the recombination time-scale
then gas behaves almost isothermally with constant ionization fraction.
For the time-scale of the sound wave larger than recombination time-scale ions begin to recombine at the end of compression.
In some astrophysical cases cooling is produced by collisional excitation of gas particles with electrons and neutrals.
 The cooling decreases with decreasing ionization fraction, leading to additional pressure increase in the element of gas.
During the rarefaction phase the pressure excess produces
work on the ambient gas and hence amplifies the wave.  In this process
some part of the energy of cosmic rays, X-rays and ultraviolet
radiation transforms into the energy of growing acoustic waves.

In this paper, we follow~\cite{dudorov99} and generalize approach of \citet{flannery} by including magnetic field into consideration.
In this case, magnetosonic modes arise. We investigate the instability of magnetosonic waves taking into account heating/cooling and ionization/recombination processes and call further this instability as a magnetic ionization-thermal instability (MITI).
We consider possible applications of MITI, such as the formation of TSAS and the generation of turbulence in the diffuse ISM.

The paper is organized in the following way.
In Section 2,  we derive
the dispersion equation for the system of magnetohydrodynamical equations using the method
of small perturbations.
In Section 3, the analytical criteria for the instability
are derived. The numerical solutions of the
dispersion equation are presented in Section 4.
Main results and their applications are discussed
in Section 5.
\section{Dispersion Equation}

Let us consider stability of  homogeneous infinite
magnetized medium with respect to small perturbations.

The behaviour of such a medium may be described
by the following set of magnetohydrodynamical (MHD) equations:
\begin{equation}
\label{eqContinuity}\frac{\partial \rho }{\partial t}+{\bf \nabla \cdot }%
\left( \rho {\bf v}\right) =0,
\end{equation}
\begin{equation}
\label{eqMotion}\frac{\partial {\bf v}}{\partial t}+\left( {\bf v\cdot }%
\nabla \right) {\bf v=}-\frac{{\bf \nabla }P}\rho+ \frac 1{4\pi \rho }\left( {\bf %
\nabla \times B}\right)\times {\bf B},
\end{equation}
\begin{equation}
\label{eqInduction}\frac{\partial {\bf B}}{\partial t}={\bf \nabla }\times
\left( {\bf v\times B}\right),
\end{equation}
\begin{equation}
\label{eqThermal}\rho \left[ \frac{\partial \varepsilon }{\partial t}+\left( {\bf %
v\cdot \nabla }\right) \varepsilon \right] +P{\bf \nabla \cdot v=-}{\cal L}
(\rho, T, x),
\end{equation}
\begin{equation}
\label{eqIonization}\frac{\partial x}{\partial t}+\left( {\bf v\cdot \nabla }%
\right) x=-{\cal R}(\rho, T, x),
\end{equation}
\begin{equation}
\label{eqState}
P=\frac{\rho k_\textrm{B} T}{\umu m_\textrm{H}},\qquad \varepsilon = \frac{P}{(\gamma -1)\rho },
\end{equation}
where
${\cal L} = \Lambda (\rho,\, T,\,x)- \Gamma (\rho,\, T,\,x) $ is the net cooling function per unit volume, $\Lambda (\rho,\, T,\,x)$ is the cooling function, $ \Gamma (\rho,\, T,\, x)$ is the heating function, $x=n_{{\rm e}}/n$ is the ionization fraction, $n_{{\rm e}}$ is the electrons concentration, $n$ is the gas density,
${\cal R}={\cal R}(\rho,\, T,\,x)$ is the net
recombination rate per atom (equal to the difference between
recombination and ionization rates),
$k_\textrm{B}$ is the Boltzmann constant,
$\umu$ is the molecular weight of the gas, and
adiabatic index $\gamma =5/3$.
All the remaining variables are used in their usual notation.

The system of equations (\ref{eqContinuity}--\ref{eqState}) takes into account non-stationary
ionization and thermal processes and allows us to investigate the effect of magnetic field on the ionization-thermal instability.
We neglect direct viscous and diffusional
processes.
 The net cooling function depends not only on the temperature
and density but also on the ionization fraction.
Equation of non-stationary
ionization (\ref{eqIonization}) contains the source term ${\cal R}(\rho,\, T,\,x)$ depending on the ionization
fraction, temperature and density.

We investigate MITI
using the method of small perturbations.
In the equilibrium state ${\cal L}(\rho,\, T,\,x)=0$ and ${\cal R}(\rho,\, T,\,x)=0$,
although the derivatives of  ${\cal L}$ and ${\cal R}$ are non-zero.
The gas is ionized by cosmic rays and is heated by
cosmic rays, X-rays and electronic photoemission from the dust grains. The
cooling  is produced by collisional excitation of fine structure levels
of the carbon ions C~\textsc{ii} by electrons and neutral hydrogen \citep{wolfire}.

We express each variable in the system of Equations~(\ref{eqContinuity}--\ref{eqState}) as a sum $f=f_\textrm{0}+f^{\prime}$, where $f_\textrm{0}$  describes the unperturbed state, and  $|f^{\prime}| \ll |f_\textrm{0}|$ is the small perturbation.
Linearising the system of Equations~(\ref{eqContinuity}--\ref{eqState}),
we obtain the following system of equations for the small perturbations
\begin{equation}
\label{eqLinContinuity}\frac{\partial \rho ^{\prime}}{\partial t}+%
\rho _{0}\nabla \cdot {\bf v} =0,
\end{equation}
\begin{equation}
\label{eqLinMotion}\frac{\partial {\bf v}^{\prime}}{\partial t}=%
-c_T^2\left(\frac{\nabla \rho ^{\prime}}{\rho _0}+\frac{\nabla T^{\prime}}{T_0}\right)+%
 \frac 1{4\pi \rho _0 }\left( {\bf %
\nabla \times B}^{\prime}\right)\times {\bf B}_0  ,
\end{equation}
\begin{equation}
\label{eqLinInduction}\frac{\partial {\bf B}^{\prime}}{\partial t}={\bf \nabla }\times
\left( {\bf v}^{\prime}{\bf \times B}_0\right),
\end{equation}
\begin{equation}
\label{eqLinThermal}
\frac{1}{T_0}\frac{\partial T^{\prime}}{\partial t}+(\gamma -1)\nabla\cdot {\bf v}^{\prime}=%
-\frac{\gamma -1}{\rho _0c_T^2}\left({\cal L}_{\rho}\rho ^{\prime}+{\cal L}_TT^{\prime}+{\cal L}_xx^{\prime}\right),
\end{equation}
\begin{equation}
\label{eqLinIonization}\frac{\partial x^{\prime}}{\partial t}=-\left({\cal R}_{\rho}\rho ^{\prime}+{\cal R}_TT^{\prime}+{\cal R}_xx^{\prime}\right),
\end{equation}
where ${\cal L}_\rho$ and ${\cal R}_{\rho}$, ${\cal L}_T$ and ${\cal R}_{T}$, ${\cal L}_x$ and ${\cal R}_{x}$ mean partial derivatives of the net cooling and net recombination functions with respect to density, temperature and ionization fraction,  respectively, $c_T=\sqrt{k_{{\rm B}}T_0/\umu m_{{\rm H}}}$ is the isothermal sound speed.

Consider the perturbations in the form
\begin{equation}
f^{\prime }=f_1\exp (i{\bf k \cdot r+}\sigma t ),\label{Eq:fp}
\end{equation}
where $f_1$ is the perturbation amplitude, ${\bf k}$ is the wave vector and
\begin{equation}
\sigma = \sigma_{{\rm R}} +i\omega,\label{sigmaReOmega}
\end{equation}
$\sigma_{{\rm R}}$ is the increment (or decrement) and $\omega$ is the frequency.
Substituting the perturbations (\ref{Eq:fp}) into
the system of Equations (\ref{eqLinContinuity}--\ref{eqLinIonization}),  we obtain the linear algebraic system
\begin{equation}
\label{eqContLin}
\sigma \frac{\rho _1}{\rho _0}+i\left({\bf k \cdot v}_1\right)=0,
\end{equation}
\begin{equation}
\label{eqMotionLin}
\sigma {\bf v}_1=-i{\bf k}c_T^2\left(\frac{\rho _1}{\rho _0}+\frac{T_1}{T_0}\right)
+\frac{i}{4\pi\rho _0}\left({\bf k} \times {\bf B_1}\right) \times{\bf B}_0,
\end{equation}
\begin{equation}
\label{eqInductionLin}
\sigma {\bf B}_1=i{\bf k} \times \left({\bf v}_1 \times {\bf B}_0\right),
\end{equation}
\begin{equation}
\label{eqThermalLin}
\sigma\frac{T_1}{T_0}+i(\gamma -1)({\bf k \cdot v}_1)=-c_T\left(k_{\rho}\frac{\rho _1}{\rho _0}+k_T\frac{T_1}{T_0}+k_x\frac{x_1}{x_0}\right).
\end{equation}
\begin{equation}
\label{eqIonizationLin}
\sigma\frac{x_1}{x_0}=-c_T\left(q_{\rho}\frac{\rho _1}{\rho _0}+q_T\frac{T_1}{T_0}+q_x\frac{x_1}{x_0}\right).
\end{equation}
We define the characteristic thermal
wave numbers in Equation~(\ref{eqThermalLin}) following \citet{field1}
$$
k_{\rho}=\frac{(\gamma -1)}{c_T^3}
\left(\frac{\partial {\cal L}}{\partial \rho}\right)_{T,x},\
k_T=\frac{(\gamma -1)T_0}{\rho_0 c_T^3}
\left(\frac{\partial {\cal L}}{\partial T}\right)_{\rho ,x},
$$

$$\
k_x=\frac{(\gamma -1)x_0}{\rho_0 c_T^3}
\left(\frac{\partial {\cal L}}{\partial x}\right)_{\rho ,T},
$$
and the characteristic ionization wave numbers in Equation~(\ref{eqIonizationLin})

$$
q_{\rho}=\frac{\rho_0}{x_0 c_T}
\left(\frac{\partial {\cal R}}{\partial \rho}\right)_{T,x},\
q_T=\frac{T_0}{x_0 c_T}
\left(\frac{\partial {\cal R}}{\partial T}\right)_{\rho ,x},\
q_x=\frac{1}{c_T}
\left(\frac{\partial {\cal R}}{\partial x}\right)_{\rho, T}.
$$

We also define following modified characteristic thermal wave numbers:

$$
\left(k_T\right)_{\rho}=k_T-\frac{q_T}{q_x}k_x,\
\left(k_{\rho}\right)_T=k_{\rho}-\frac{q_{\rho}}{q_x}k_x,
$$
$$
\left(k_T\right)_P=\left(k_T\right)_{\rho}-\left(k_{\rho}\right)_T.
$$

Resolving equations (\ref{eqContLin}, \ref{eqThermalLin}, \ref{eqIonizationLin}) with respect to $\rho_1/\rho _0$, $T_1/T_0$ and $x_1/x_0$, we reduce Equation (\ref{eqMotionLin}) to
\begin{equation}
\label{eqMotionLin1}
\sigma {\bf v}_1+
\frac{\left(
{\bf k \cdot v}_1 \right)}{\sigma}
\widetilde{\gamma }c^2_T{\bf k}
=\frac{i}{4\pi\rho _0}\left({\bf k \cdot B}_1\right){\bf B}_0,
\end{equation}
where
\begin{equation}
\label{eqEffGamma}
\widetilde{\gamma }=\frac
        {
          \gamma \sigma ^2+c_T\left(
          k_T-k_{\rho}+\gamma q_x\right)\sigma +
          c_T^2q_x(k_T)_P
        }
        {
          \sigma ^2+c_T\left(k_T+q_x\right)\sigma +
          c_T^2q_x(k_T)_{\rho}
        }
\end{equation}
plays the role of `effective' adiabatic index.
Expanding the double vector products in Equation (\ref{eqInductionLin})
and substituting ${\bf B}_1$ from Equation (\ref{eqInductionLin})
into Equation (\ref{eqMotionLin1}),
we obtain
the following
characteristic vector equation for the velocity perturbation:
\begin{equation}
\label{eqMainDispersion}
\begin{array}{c}
\displaystyle
\sigma ^2 {\bf v}_1 +
\left( {\bf k \cdot v}_1 \right)
\left[ \widetilde{\gamma }c^2_T{\bf k} +
       {\bf v}_{{\rm A}} \times \left({\bf k} \times {\bf v}_{{\rm A}}\right)
\right] - \\
\displaystyle
\left({\bf k \cdot v}_{{\rm A}}\right)
\left[{\bf v}_{{\rm A}} \times \left({\bf k} \times {\bf v}_1\right)\right]=0,
\end{array}
\end{equation}
where
${\bf v}_{{\rm A}}={\bf B}_0/\sqrt{4\pi\rho _0}$ is the Alfv\`en speed.

Equation (\ref{eqMainDispersion}) describes the waves with various
relative orientations of vectors ${\bf k,v}_1,$ and ${\bf v}_{{\rm A}}$.
The effects of thermal and magnetic pressures in this equation
are described by the second term, while the third one describes the
magnetic tension effects.

Equation
(\ref{eqMainDispersion}) reduces to the dispersion equation
for the Alfv\`en waves in the case ${\bf v}_1\perp {\bf k}$,
$$
\sigma ^2+\left({\bf k \cdot v}_{{\rm A}}\right)^2=0.
$$
This equation does not include the effects
of ionization-recombination and heating-cooling processes, as
the Alfv\`en
waves are incompressible ones in linear approximation. We do not consider them in the following.

Making separately vector and scalar products of ${\bf k}$ with
Equation (\ref{eqMainDispersion}), we obtain a homogeneous system of
linear algebraic equations with unknowns $\left({\bf k} \times {\bf v}_1\right)$ and
$\left({\bf k \cdot v}_1\right)$.
Resolving this system, we obtain a
dispersion equation for the magnetosonic and dynamical modes:
\begin{equation}
\label{eqMagnetosonic}
\sigma ^4+\sigma ^2\left(\widetilde{\gamma}c^2_{T}+v^2_{{\rm A}}\right)k^2+
\widetilde{\gamma}c^2_{T}v^2_{{\rm A}}k^4\cos^2\theta=0,
\end{equation}
where $\theta $
is the angle between the wave vector and the unperturbed magnetic field.

Substituting Expression (\ref{eqEffGamma}) into Equation (\ref{eqMagnetosonic}), we obtain a
dispersion equation of the 6-th order
\begin{equation}
\label{eqDispersion}
\begin{split}
\sigma^6&+c_T\left(k_T+q_x\right)\sigma ^5+
c_T^2\left[\left( \gamma +A\right) k^2+
q_x\left(k_T\right)_{\rho}\right]\sigma ^4+ \\
&
c_T^3 k^2\left[ \left(1+A\right)
k_T -k_{\rho}+\left(\gamma +
A\right) q_x\right]\sigma ^3+\\
&
c_T^4 k^2\left[
A\gamma k^2\cos^2 \theta +
q_x
\left(k_T\right)_{\rho}\left(1+A\right)
-q_x \left(k_{\rho}\right)_T
\right]\sigma ^2+\\
&
A c_T^5 k^4\cos^2 \theta
\left(k_T-k_{\rho}+\gamma q_x\right)\sigma +\\
&
Ac_T^6 q_x\left(k_T\right)_P k^4\cos^2\theta=0,
\end{split}
\end{equation}
where
\begin{equation}
A=v_{{\rm A}}^2/c_T^2=2/\beta
\label{alnum}
\end{equation}
is the Alfv\`en number, $\beta =8\pi P/B^2$ is the plasma beta.

\section{Instability Criteria}

\subsection{Asymptotic analysis}
\label{Sec:Asym}
The dispersion equation (\ref{eqDispersion}) is the polynomial of the
sixth order and it cannot be solved analytically in general.
We can find out all of its solutions in the limiting cases
of small and large wave numbers by means of asymptotic analysis
following \citet{heyvaerts}.

\subsubsection{Small wavelength limit}
\label{Sec:klarge}
Let us consider the case of small wavelengths.
The solutions of Equation (\ref{eqDispersion}) include
two slow magnetosonic modes

\begin{equation}
\label{asympslowl0}
\sigma _{{\rm s} \pm} = -\frac{c_T\left(v_{{\rm f}}^2-\gamma c_T^2\right)}
{2\gamma \left(v_{{\rm f}}^2-v_{{\rm s}}^2\right)}\left[k_T\left(\gamma - 1\right)+k_\rho\right]
\pm ikv_{{\rm s}},
\end{equation}

\noindent
two fast magnetosonic modes

\begin{equation}
\label{asympfastl0}
\sigma _{{\rm f} \pm} = -\frac{c_T\left(\gamma c_T^2-v_{{\rm s}}^2\right)}
{2\gamma \left(v_{{\rm f}}^2-v_{{\rm s}}^2\right)}\left[k_T\left(\gamma - 1\right)+k_\rho\right]
\pm ikv_{{\rm f}},
\end{equation}

\noindent
and the pair of dynamical modes
\begin{equation}
\label{asympthrl0}
\begin{split}
\sigma _{P \pm}=&\frac{c_T}{2\gamma}
\left[-\left(k_T-k_{\rho}+\gamma q_x\right) \right]\pm
\\ &
\frac{c_T}{2\gamma}\sqrt{\left(k_T-k_{\rho}+\gamma q_x\right)^2-4\gamma q_x\left(k_T\right)_P},
\end{split}
\end{equation}
where
\begin{equation}
\label{eqVfVs}
v_{{\rm f},{\rm s}}=\sqrt{\frac{1}{2}\left[\gamma c_T^2+v_{{\rm A}}^2\pm
\sqrt{\left(\gamma c_T^2+v_{{\rm A}}^2\right)^2-4\gamma c_T^2v_{{\rm A}}^2\cos^2\theta}\right]}
\end{equation}
are the fast ($v_{{\rm f}}$) and slow ($v_{{\rm s}}$) magnetosonic speeds. The fast  and slow waves correspond to `$+$' and `$-$' signs under the square root in Equation~(\ref{eqVfVs}), respectively.

\citet{heyvaerts} found the solutions similar to (\ref{asympslowl0}--\ref{asympfastl0})
taking into account
Joule and thermal conductivity terms. The magnetosonic modes (\ref{asympslowl0}--\ref{asympfastl0}) are unstable if
\begin{equation}
\label{neqisentr}
k_T\left(\gamma - 1\right)+k_\rho<0 .
\end{equation}
Inequality (\ref{neqisentr}) is the criterion for the isentropic instability \citep{field1}. It is not satisfied in the conditions of diffuse H~\textsc{i} clouds that we consider in this paper
\citep{oppenheimer}.

Equations (\ref{asympslowl0}--\ref{asympfastl0}) show that
the ionization and recombination processes do not affect the magnetosonic
modes at  small wavelengths.

Modes (\ref{asympthrl0})
are unstable if one of inequalities
\begin{equation}
\label{crIsobaricTI}
k_T-k_{\rho}+\gamma q_x <0,
\end{equation}
\begin{equation}
\label{crIsobaricITI}
\left(k_T\right)_P<0,
\end{equation}
or both of them are satisfied.
Inequality (\ref{crIsobaricTI}) corresponds to
the \citeauthor{field1}'s (\citeyear{field1}) criterion
for the isobaric thermal instability, but it takes into account the
stabilization due to recombinations.
Inequality (\ref{crIsobaricITI}) corresponds to the thermal-reactive
instability criterion \citep{yoneyama,corbelli}.
When inequality (\ref{crIsobaricTI}) is satisfied and (\ref{crIsobaricITI}) is not, then Equation~(\ref{eqDispersion})
has two unstable roots instead of one root of the
dispersion equation of \citet{field1}. This effect occurs due to an additional degree of freedom that appears from the ionization equation.

The pair of modes (\ref{asympthrl0}) can transform to oscillatory ones.
It depends on the sign of the expression under the square root in
Equation (\ref{asympthrl0}). If
$$
\left(k_T-k_{\rho}+\gamma q_x\right)^2-4\gamma q_x\left(k_T\right)_P<0
$$
then Equation (\ref{asympthrl0}) becomes
\begin{equation}
\label{asympthrl0osc}
\begin{split}
\sigma _{P\pm}=&
-\frac{c_T}{2\gamma}\left(k_T-k_{\rho}+\gamma q_x\right) \pm \\ & \quad
i\frac{c_T}{2\gamma}\sqrt{\left|
\left(k_T-k_{\rho}+\gamma q_x\right)^2-4\gamma q_x\left(k_T\right)_P
\right|}.
\end{split}
\end{equation}
Modes (\ref{asympthrl0osc}) develop as the standing waves, as their frequency
does not depend on the
wave number and their group wave speed equals zero.
\textcolor{red}{}

\subsubsection{Large wavelength limit}
\label{Sec:ksmall}
In the long wavelength limit, the solutions of Equation (\ref{eqDispersion}) are two slow magnetosonic modes
\begin{equation}
\label{asympslowlinf}
\sigma _{{\rm s} \pm} = -\frac{c_T}{2}\frac{Q_{{\rm s}}}
{q_x\left(k_T\right)_{\rho}\left(k_T\right)_P}
\xi k^2 \pm ikv_{{\rm s}},
\end{equation}

\noindent
two fast magnetosonic modes
\begin{equation}
\label{asympfastlinf}
\sigma _{{\rm f} \pm} = -\frac{c_T}{2}\frac{Q_{{\rm f}}}
{q_x\left(k_T\right)_{\rho}\left(k_T\right)_P}
\xi k^2 \pm ikv_{{\rm f}},
\end{equation}

\noindent
and two dynamical modes
\begin{equation}
\label{asympthrlinf}
\sigma _{\rho \pm} = \frac{c_T}{2}
\left[
  -\left(
     k_T+q_x
   \right)
\pm
\sqrt{
\left(
   k_T+q_x
\right)^2
-4q_x
\left(
k_T
\right)_{\rho } }
\right],
\end{equation}
where
\begin{equation}
\label{oscITI}
\xi =
        q_x
        \left[
          \left(
            \gamma -1
          \right)
          \left(
            k_T
          \right)_{\rho}+
          \left(
            k_{\rho}
          \right)_T
        \right]
        +k_T
        \left(
          k_{\rho}
        \right)_T-
        k_{\rho}
        \left(
          k_T
        \right)_{\rho},
\end{equation}
\begin{equation}
\label{QsQf}
Q_{{\rm s}}=\frac{v_{{\rm s}}^2\left(v_{{\rm f}}^2-\gamma _0 c_T^2\right)}{c_T^2\left(v_{{\rm f}}^2-v_{{\rm s}}^2\right)},
Q_{{\rm f}}=\frac{v_{{\rm f}}^2\left(\gamma _0 c_T^2-v_{{\rm s}}^2\right)}{c_T^2\left(v_{{\rm f}}^2-v_{{\rm s}}^2\right)},
\end{equation}
and
\begin{equation}
\label{gammaprime}
\gamma _0 = \frac{\left(k_T\right)_P}{\left(k_T\right)_{\rho}}
\end{equation}
is the effective adiabatic index
in the small frequency limit.
The wave speeds of the fast and slow waves are
expressed by Equation~(\ref{eqVfVs}) with $\gamma _0$ substituted instead
of $\gamma$.

The fast and slow magnetosonic modes (\ref{asympslowlinf}--\ref{asympfastlinf}) are
 unstable if
\begin{equation}
\label{crOscITI1}
\frac{\xi}{q_x(k_T)_{\rho}(k_T)_P}<0.
\end{equation}
In the isobarically and isochorically stable medium
condition (\ref{crOscITI1}) becomes
\begin{equation}
\label{crOscITI}
\xi <0.
\end{equation}
This is a criterion of MITI. Expressions (\ref{oscITI}) and (\ref{crOscITI1}) show that this criterion does not include magnetic field, and it is the same as for the ionization-coupled acoustic instability of \citet{flannery}.

\begin{figure}
	\includegraphics[width=\columnwidth]{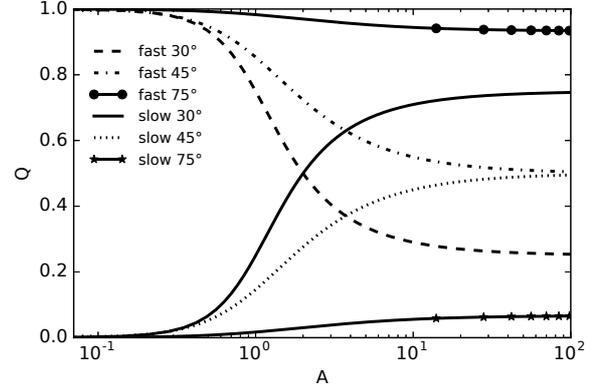}
   \caption{The dependencies of the dimensionless factors $Q_{{\rm f}}$ (dashed line, dot-dashed line and solid line with rounds) and $Q_{{\rm s}}$ (solid line, dotted line and solid line with stars) on the Alfv\`en number $A$. The curves are plotted for $\theta= 30^{\circ}, 45^{\circ}, 75^{\circ}$ (see legend) and adiabatic index $\gamma_0=1$.}
    \label{fig:Q}
\end{figure}

The growth rates of the slow and fast magnetosonic modes (\ref{asympslowlinf}), (\ref{asympfastlinf}) are determined by the
dimensionless factors $Q_{{\rm s}}$ and $Q_{{\rm f}}$ that depend on the Alfv\`en number (\ref{alnum})
through $v_{{\rm s}}$ and $v_{{\rm f}}$.
If $Q_{{\rm s}}/Q_{{\rm f}}>1$ then the slow
mode has faster growth rate than the fast one.
In Fig.~\ref{fig:Q}~, we plot the dependencies of $Q_{{\rm s}}$ and $Q_{{\rm f}}$ on $A$
for
various values of angle $\theta$. Fig.~\ref{fig:Q} shows that the fast mode dominates
the slow one at any $\theta$ in the case of weak magnetic field $A\ll 1$ (or $\beta\gg 1$ according to Equation (\ref{alnum})).
The slow mode dominates the fast one at  $\theta < \upi /4$  in the case strong magnetic field $A\gg 1$ ($\beta\ll 1$).

Modes (\ref{asympthrlinf}) are unstable if
\begin{equation}
\label{crIsochoricTI}
k_T+q_x<0
\end{equation}
or
\begin{equation}
\label{crIsochoricITI}
\left(k_T\right)_{\rho}<0.
\end{equation}
Criterion (\ref{crIsochoricTI}) corresponds to the isochoric
thermal instability of \cite{field1} with stabilization due recombinations.
Inequality (\ref{crIsochoricITI}) describes the thermal-reactive instability in the
large wavelength limit \citep{corbelli}.


\subsection{Critical wave numbers and frequencies}

The asymptotic analysis carried out in Section~\ref{Sec:Asym} shows that the  magnetosonic modes
in the isobarically, isochorically and isentropically stable medium are unstable only if
condition (\ref{crOscITI}) for $\xi$ from (\ref{oscITI}) is satisfied.
In this case, the magnetosonic
mode is unstable only at the wave numbers less than some critical one.
In order to determine the critical wave numbers and corresponding
frequencies for the slow and fast magnetosonic modes, we represent $\sigma$ in the form (\ref{sigmaReOmega}).
The critical wave numbers and frequencies correspond to the case $\sigma _{{\rm R}}=0$.
We substitute $\sigma=i\omega$ into the dispersion equation (\ref{eqDispersion}) and
obtain the following critical wave number and corresponding critical frequency:

\begin{equation}
\label{kcrfs2}
k_{{\rm cr}}^2=-\frac{q_xc_T^2\xi}{u_{{\rm f,s}}^2\left[k_T\left(\gamma - 1\right)+k_\rho\right]},
\end{equation}
\begin{equation}
\label{wfs2}
\omega _{{\rm cr}}^2=-\frac{q_xc_T^2\xi}{k_T\left(\gamma - 1\right)+k_\rho},
\end{equation}
where $u_{{\rm f,s}}$ are the fast and slow magnetosonic speeds determined according to Equation (\ref{eqVfVs}) with $\gamma _{{\rm cr}}$ substituted instead of $\gamma$.
And
$$
\gamma _{{\rm cr}} =\frac{k_T-k_{\rho}+\gamma q_x}{k_T+q_x}
$$
is the effective adiabatic index for the waves with the critical frequency
$\omega _{{\rm cr}}$. It should be noted that frequency $\omega _{{\rm cr}}$ in Equation~(\ref{wfs2})
is the same for the fast and slow magnetosonic waves.

Formulae~(\ref{kcrfs2}-\ref{wfs2}) determine the regions of the parameter space
in which the magnetosonic waves can be amplified, $k<k_{{\rm cr}}$ or $\omega<\omega_{{\rm cr}}$.

\section{Numerical solution of the dispersion equation}

In this section, we  solve numerically the dispersion equation (\ref{eqDispersion}) and analyse its solutions. We apply the Bairstow's method to find out the roots.
This method uses Newton-Raphson iterations to extract the complex roots of
a polynomial by pairs.

\begin{figure}
	\includegraphics[width=\columnwidth]{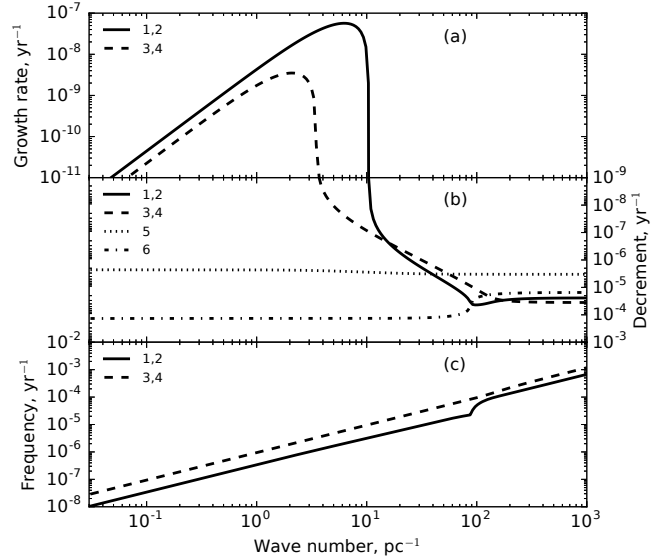}
   \caption{Two slow magnetosonic waves (solid lines 1,2), two fast magnetosonic waves (dashed lines 3,4) and pair of stable  modes
   with $\omega=0$
   (dotted and dot-dashed lines 5,6)
   are solutions of the dispersion equation (\ref{eqDispersion}) for $T = 70$~K and $\theta = \upi /6$. Panel (a): the growth rates of unstable modes ($\sigma_{{\rm R}}>0$). Panel (b): the decrements of stable modes ($\sigma_{{\rm R}}<0$). Panel (c): the frequencies ($\omega$) on the wave number.}
   \label{fig:7030}
\end{figure}

\begin{figure}
\includegraphics[width=\columnwidth]{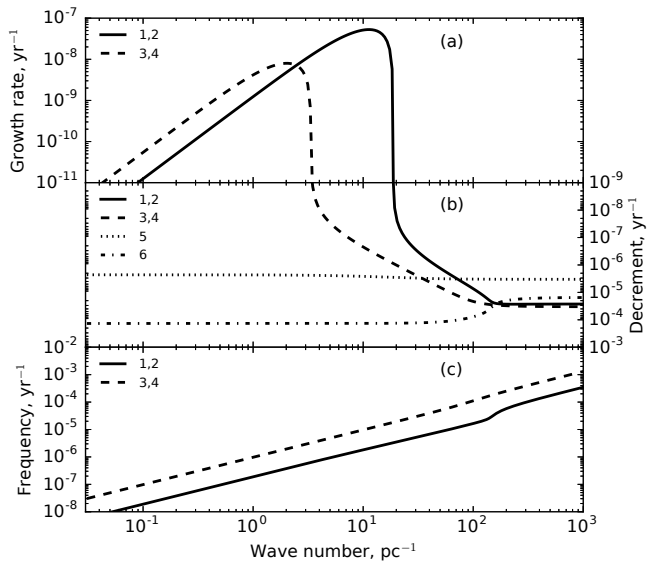}

   \caption{Same as in Fig.~\ref{fig:7030}, but for $T = 70$~K and $\theta = \upi /3$.}
    \label{fig:7060}
\end{figure}

\begin{figure}
	\includegraphics[width=\columnwidth]{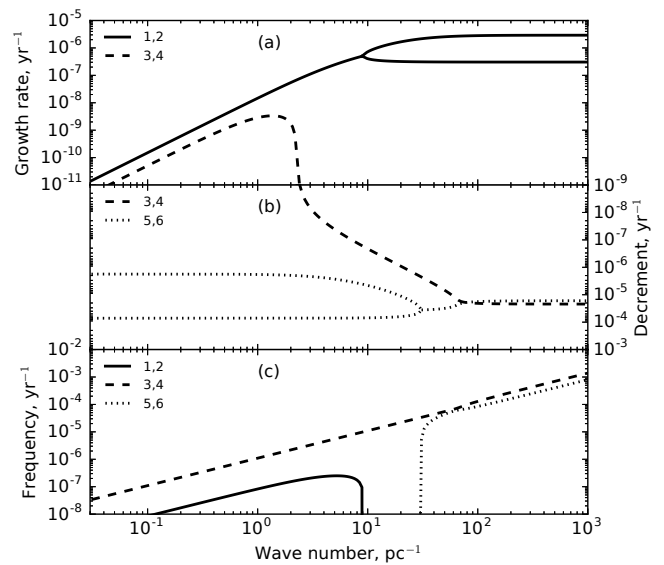}
   \caption{Same as in Fig.~\ref{fig:7030}, but for $T = 95$~K and $\theta = \upi /6$.
   Here two unstable SMSW (solid lines 1,2) transform into the pair of unstable ionization-thermal modes,
   and pair of stable modes with $\omega=0$ (dotted lines 5,6) transform into the pair of stable SMSW
   towards larger wave numbers.}
    \label{fig:9530}
\end{figure}

\begin{figure}
	\includegraphics[width=\columnwidth]{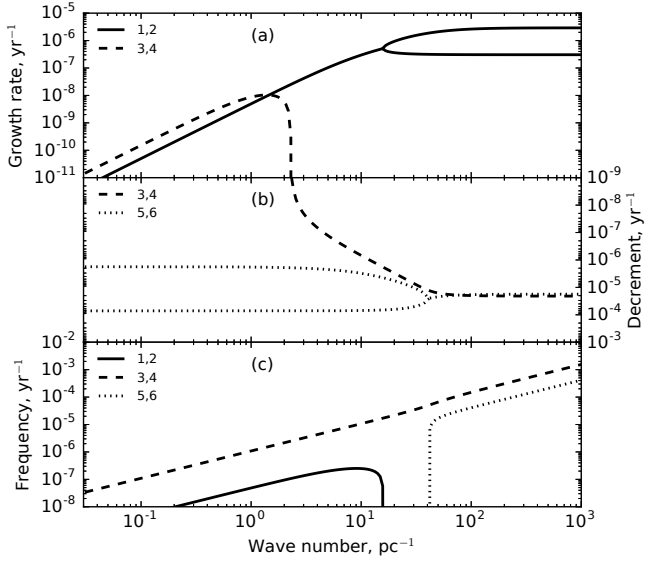}
   \caption{Same as in Fig.~\ref{fig:9530}, but for $T = 95$~K and $\theta = \upi /3$.}
    \label{fig:9560}
\end{figure}

\begin{figure}
	\includegraphics[width=\columnwidth]{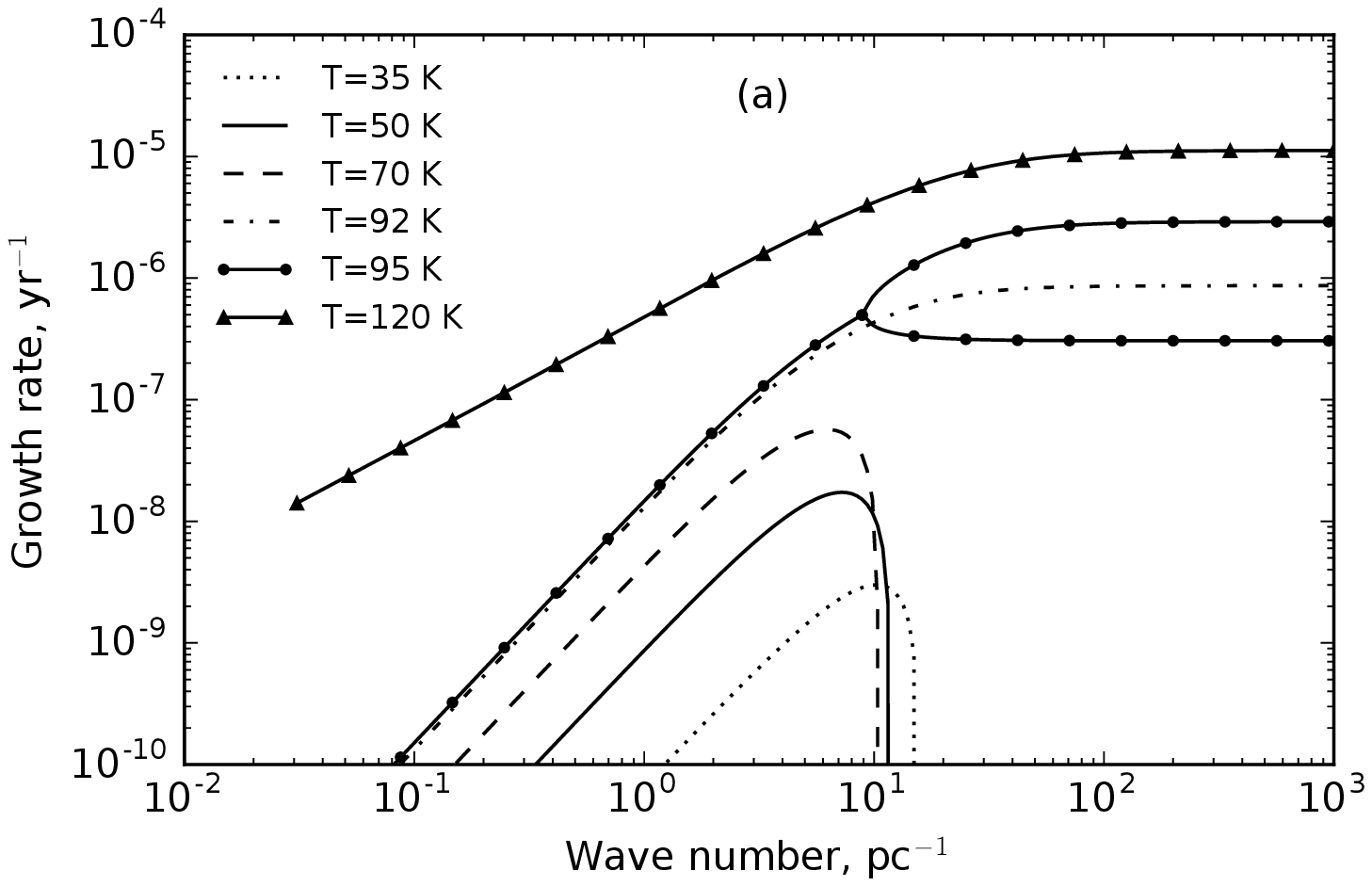}
    \includegraphics[width=\columnwidth]{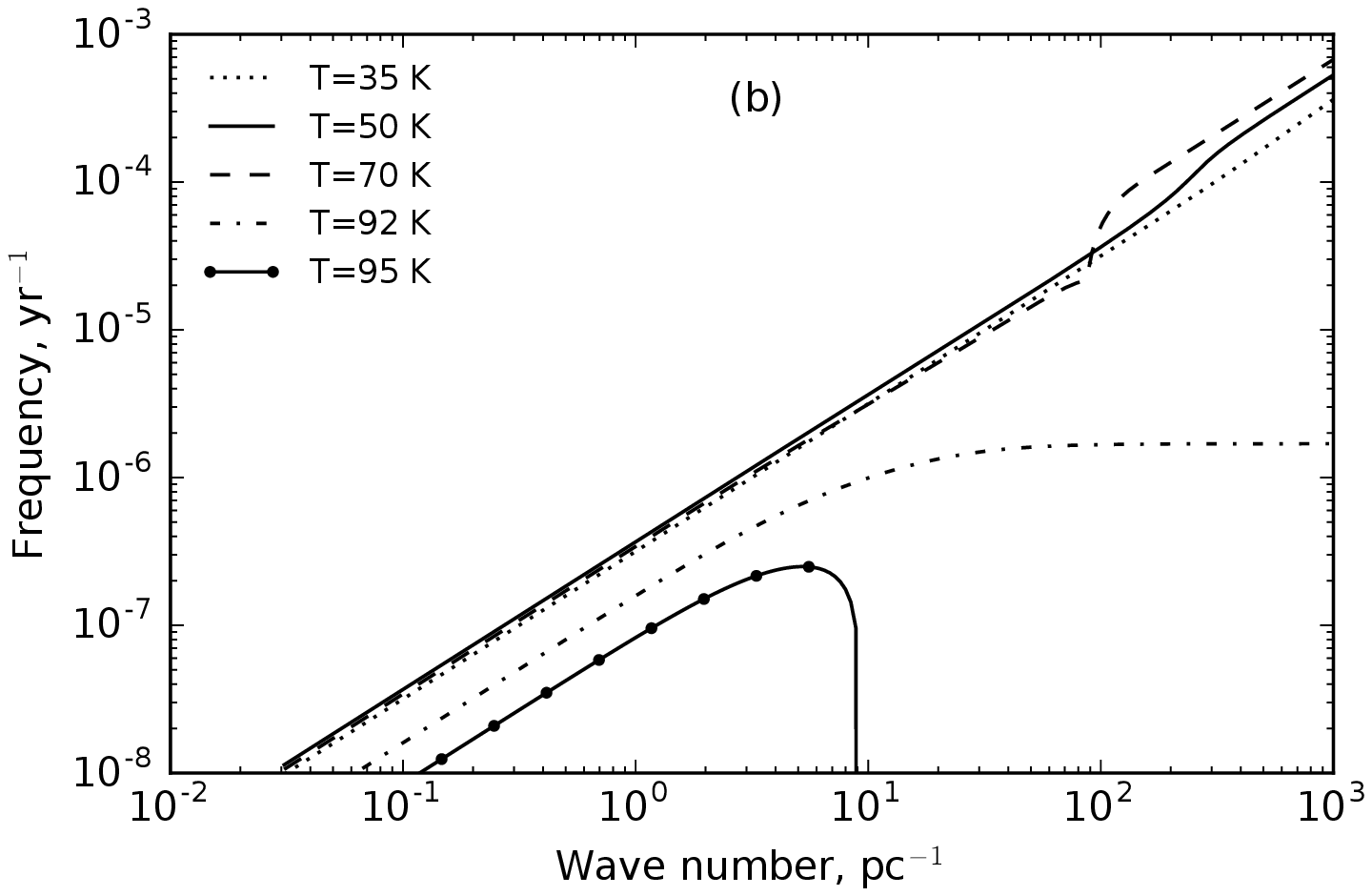}
   \caption{The dependencies of the growth rates $\sigma_{{\rm R}}$ (panel (a)) and the frequencies $\omega$ (panel (b)) of the unstable slow magnetosonic modes and dynamical modes on the wave number for various values of temperature
in the range (35-120~K).  The angle between magnetic field $\bf B$ and wave vector $\bf k$ $\theta = \upi /6$.}
    \label{fig:slowdyn}
\end{figure}

\begin{figure}
	\includegraphics[width=\columnwidth]{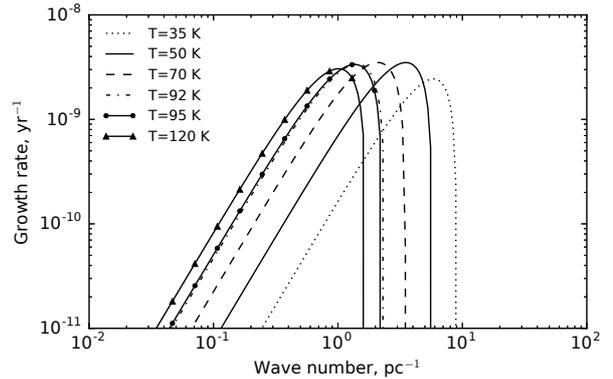}
   \caption{The dependencies of the growth rates $\sigma$ of the unstable fast
magnetosonic modes on the wave number for different values of temperature
in the range (35-120~K) for the angle between magnetic field $\bf B$ and wave vector $\bf k$ $\theta = \upi /6$.}
    \label{fig:fast}
\end{figure}

\begin{figure}
	\includegraphics[width=\columnwidth]{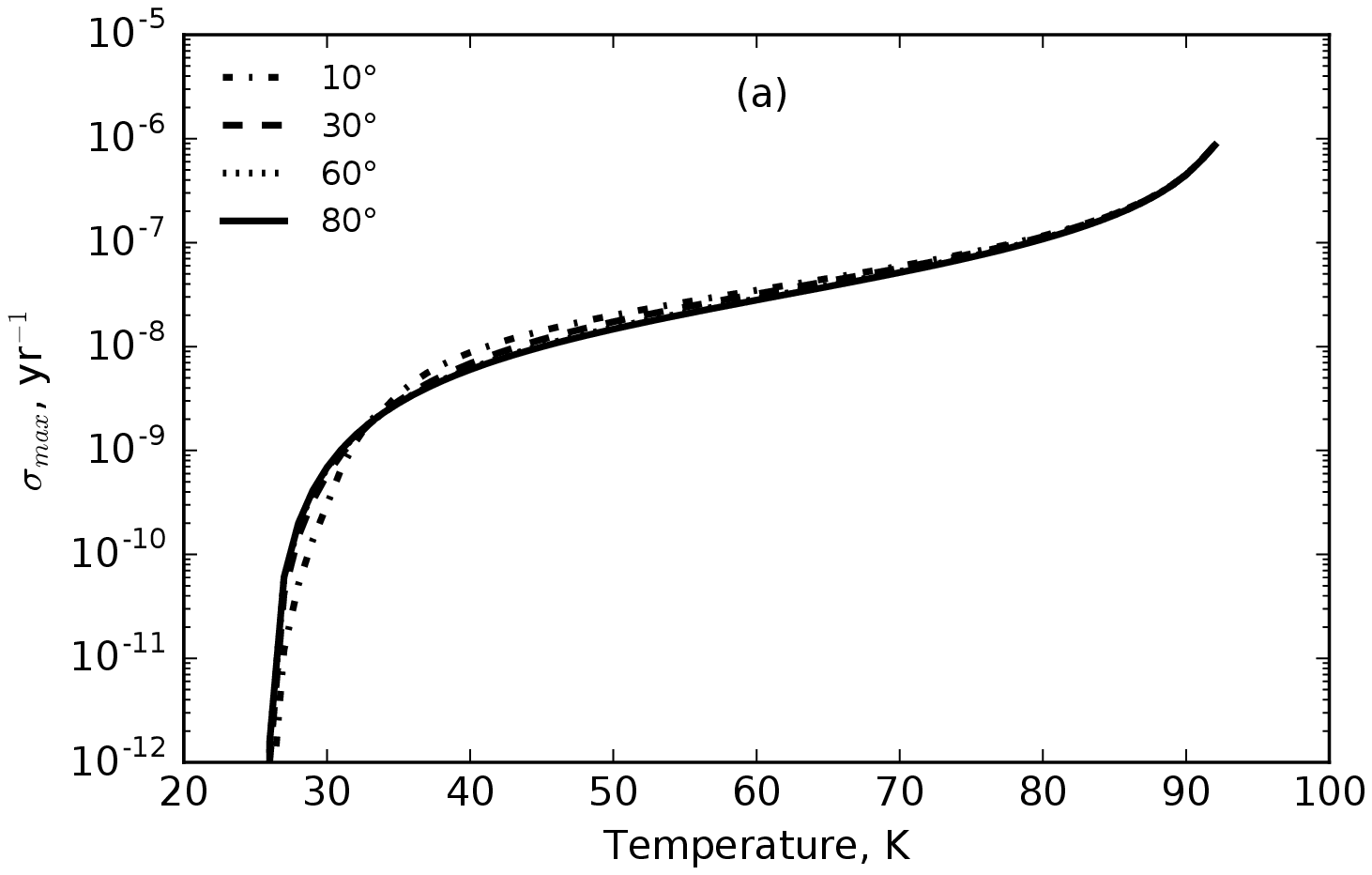}
    \includegraphics[width=\columnwidth]{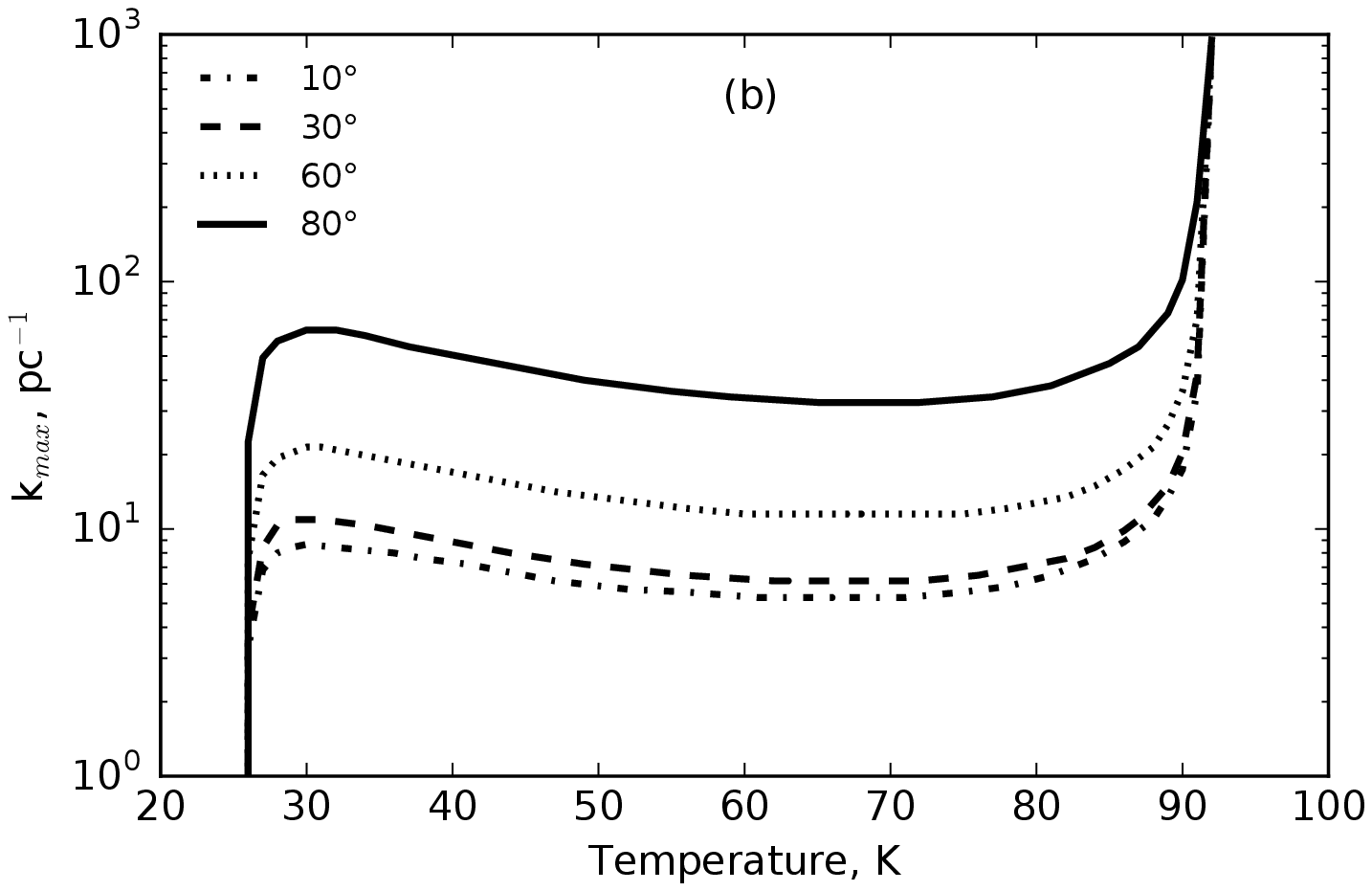}
   \caption{The dependencies of the maximum growth rate (a) and the corresponding wave number (b)
of the unstable slow magnetosonic modes on temperature for different values of the angle $\theta$.}
    \label{fig:slowmax}
\end{figure}

The results of the solution of the dispersion equation (\ref{eqDispersion})
are shown
in Figs.~\ref{fig:7030}--\ref{fig:slowmax} for the mean Galactic values of magnetic field  $B=2\cdot 10^{-6}$~G,
ionization rate of cosmic rays $\xi _\textrm{CR} = 10^{-17}$~s$^{-1}$ and heating rate $h= 5\cdot 10^{-26}$~erg~s$^{-1}$~\citep{spitzer68}.

The heating function can be expressed as $\Gamma(\rho)=nh$.
The cooling function is calculated according
to \citet{wolfire}
\begin{equation}
\label{CoolingFunc}
\begin{split}
\Lambda (\rho,\, T,\,x)=2.54 \cdot 10^{-14}X_{\textrm C}f_{\textrm{C \textsc{ii}}}\left[\gamma^{\textrm H_0}\left(1-x\right)+\gamma^ex\right]n^2\times&\\
 \textrm{exp}\left(-\frac{92}{T}\right)\
\textrm{erg s$^{-1}$ cm$^{-3}$,}
\end{split}
\end{equation}
where the carbon abundance $X_{\textrm C}=3\cdot 10^{-4} $ and fraction $f_{\textrm{C \textsc{ii}}}=1$. Values of the collisional de-excitation rate coefficients for collisions with neutral hydrogen $\gamma^{\textrm H_0}=8.86\cdot 10^{-10}$ cm$^3$ s$^{-1}$ and electrons $\gamma^e=\gamma^e(T)$ are taken from \citet{wolfire}.
Equation (\ref{CoolingFunc}) includes cooling only due the collisional excitations of fine structure levels C~\textsc{ii} with hydrogen atoms and electrons.

For the net recombination rate ${\cal R}(\rho,\, T,\,x)$ we take
\begin{equation}
\label{IonizationFunc}
{\cal R}(\rho,\, T,\,x)=x^2n\alpha(T)-\xi _\textrm{CR}(1-x),
\end{equation}
where $\alpha(T)=4.1 \cdot 10^{-12} \left(T/10^2\right)^{-0.6}$ cm$^3$ s$^{-1}$ is the rate of radiative recombinations taken from \citet{flannery}.

The density and ionization fraction are calculated for given temperature using equations (\ref{CoolingFunc}--\ref{IonizationFunc}) in the equilibrium state ${\cal L}(\rho,\, T,\,x)=0$ and ${\cal R}(\rho,\, T,\,x)=0$. For temperatures $T=30-100$~K we obtain $n\sim 150-15$ cm$^{-3}$ and $x\sim 10^{-5} - 10^{-4}$  that correspond to diffuse clouds. Corresponding plasma beta lies in range from 4 to 1.4.

For example, the characteristic wave numbers are $k_\rho=159.1$ pc$^{-1}$, $k_T=199.0$ pc$^{-1}$,
$k_x=20.2$ pc$^{-1}$,  $q_\rho=1.6$ pc$^{-1}$, $q_T=-1.0$ pc$^{-1}$ and $q_x=3.2$ pc$^{-1}$ at $T=70$ K.
For these wave numbers criterion (\ref{crOscITI}) is satisfied.

In Fig.~\ref{fig:7030}--\ref{fig:9560}, we show the solutions for various temperatures (70 and 95~K) and angles $\theta=\upi /6$, $\upi /3$.
The a-panels of each figure
show the growth rates of the unstable modes ($\sigma_{{\rm R}}>0$); the b-panels show the absolute values of the decrements
of the decaying modes ($\sigma_{{\rm R}}<0$) and the c-panels show the frequencies ($\omega$).

In Fig.~\ref{fig:7030},  we depict the solutions of Equation (\ref{eqDispersion}) for $T=70$~K and $\theta =\upi /6$. In this case, curves 1 and 2 describe slow magnetosonic waves (SMSW) propogating in the opposite directions. Curves 3 and 4 correspond to fast magentosonic waves (FMSW) propogating in the opposite directions. And curves 5, 6 are the pair of the stable isobarical modes. Growth rates of SMSW and FMSW proportional to $k^2$ for small wave numbers according to Equations (\ref{asympslowlinf}, \ref{asympfastlinf}). The growth rate of SMSW is positive for the wave numbers $k<10$~pc$^{-1}$.
Fig.~\ref{fig:7030} shows that the growth rate of SMSW increases with $k$ from $ 10^{-11}$~yr$^{-1}$ at $k\approx 0.04$~pc$^{-1}$ to maximum $5\cdot 10^{-8}$~yr$^{-1}$ at $k\approx 5$~pc$^{-1}$ and then rapidly goes to zero at $k=k_{{\rm cr}}^{{\rm SMSW}}\approx 10$~pc$^{-1}$.
At $0<k<0.03$~pc$^{-1}$ SMSW and FMSW are unstable but their growth rates are less than $10^{-11}$~yr$^{-1}$.
SMSW are stable at $k>k_{{\rm cr}}^{{\rm SMSW}}$. The decrement of SMSW increases from 0 to $10^{-5}$~yr$^{-1}$ in the range $k=[k_{{\rm cr}}^{{\rm SMSW}},\,100]$~pc$^{-1}$ and remains nearly constant at larger $k$. The growth rate and decrement of FMSW depend on $k$ in similar way as for SMSW. FMSW are unstable at  $k<k_{{\rm cr}}^{{\rm FMSW}}\approx 4$~pc$^{-1}$ and has maximum growth rate $\approx 3\cdot 10^{-9}$~yr$^{-1}$.

As it was shown with the help of asymptotic analysis (see Section~\ref{Sec:Asym}), the ionization and recombination processes do not affect the instability of the magnetosonic waves for the short wavelength perturbations
(Section~\ref{Sec:klarge}).
This instability correspond to the  isentropic thermal instability (\ref{neqisentr}) and it does not develop under the conditions considered in our paper.

The period of a magnetosonic wave equals $t_{\textrm{0}}=\lambda/(c_{{\rm s}}^2 + v_{{\rm A}}^2)^{1/2}$,
where $\lambda$ is the wavelength and $c_{{\rm s}}$ is the sound speed.
The characteristic ionization time-scale equals $t_{\textrm{ion}}=1/\xi _{\textrm{CR}}\sim 10^9$ yrs for $\xi _{\textrm{CR}}=10^{-17}$ s$^{-1}$.
The period of the magnetosonic wave with small
wavelength is much less than the ionization time.
The ionization fraction  does not change significantly over
the period of the wave, and the wave amplification mechanism does not work. Hence, the  magnetosonic waves decay at small $\lambda$.

At small $k$ magnetosonic waves are unstable but growth rates could be too low for instability to develop.
In this case the characteristic time-scale of a magnetosonic wave is much more than the cooling and recombination time-scales. The gas evolves to the ionization-thermal equilibrium, and cools too fast avoiding the rarefaction phase.

SMSW modes have
ten times faster maximum growth rate
than FMSW, and SMSW are unstable in wider range of wavelengths under considered conditions. The decrements of SMSW and FMSW are
almost equal to each other at large wave numbers.
The frequencies of SMSW and FMSW increase with $k$ according to a power law from $10^{-8}-10^{-7}$~yr$^{-1}$ at $k=0.1$~pc$^{-1}$ to $10^{-3}$~yr$^{-1}$ at $k=1000$~pc$^{-1}$.

Fig.~\ref{fig:7060} shows that general behaviour of the solutions for $\theta=\upi /3$ is similar to the one depicted in Fig.~\ref{fig:7030}. Maximum growth rate of SMSW shifts towards larger $k$, while maximum growth rate of FMSW moves towards smaller $k$, as compared to the case with $\theta=\upi /6$. The FMSW modes slightly dominate the SMSW ones at small wave numbers, $k\la k_{{\rm cr}}^{{\rm FMSW}}\approx 3$~pc$^{-1}$. At higher wave numbers, FMSW decays, while the SMSW mode is unstable up to $k\la k_{{\rm cr}}^{{\rm SMSW}}\approx 20$~pc$^{-1}$.  The
slight dominance of SMSW over FMSW
for small wave numbers
at $\theta =\upi /6$ and conversely at $\theta =\upi /3$ confirms the results of asymptotic analysis (see Section~\ref{Sec:ksmall}). The maximum growth rates are of the same order as in the case $\theta =\upi /6$.

We pay attention to SMSW as a mode with the largest growth rate and define the right bound for instability as $k_{cr}=k_{{\rm cr}}^{{\rm SMSW}}$.
In Table~\ref{tab:k}, we listed values of $k_{{\rm cr}}$ for the slow magnetosonic waves as a function of temperature (different rows) and angle $\theta$ (different columns) calculated using analytic formulae~(\ref{kcrfs2}). The values of $k_{{\rm cr}}$ correspond to the right boundary of the instability domain for the magnetosonic waves.
Table~\ref{tab:k} shows that $k_{{\rm cr}}$ increases with $\theta$. The dependence of $k_{{\rm cr}}$ on $T$ is non-monotonic and has minimum at
$T \approx 65$~K for $\theta=10^\circ-80^\circ$. The values of the critical wave numbers $k_{{\rm cr}}^{{\rm SMSW}}$ found in the numerical solution of the dispersion equation (see Fig.~\ref{fig:7030}, \ref{fig:7060}) agree with the values in Table~\ref{tab:k}. The values of $k_{{\rm cr}}$ lie between $\approx 10$ and $\approx 100$~pc$^{-1}$, that corresponds to the waves with $\lambda\approx[0.01,\,0.1]$~pc.

Figs.~\ref{fig:9530}--\ref{fig:9560} show the solutions of the dispersion equation for the case $T=95$~K. The transition to the isobaric thermal instability occurs near this temperature.

At this temperature, the solutions 1 and 2 correspond to the unstable
slow magnetosonic modes at $k< 8.8\, {\rm pc}^{-1}$ $(\theta =\upi /6)$
and $k<15.6\, {\rm pc}^{-1}$ $(\theta = \upi /3)$. At larger wave numbers,
these solutions describe the unstable ionization-thermal modes. Solutions 5 and 6 (dotted lines) correspond to two  decaying modes at $k< 30.6\, {\rm pc}^{-1}$ $(\theta = \upi /6)$ and $k< 41.8\, {\rm pc}^{-1}$ $(\theta = \upi /3)$ that transform into
the pair of decaying slow magnetosonic modes at larger wave numbers. There is a `gap' of wave numbers at which the slow magnetosonic
waves do not propagate $(8.8< k< 30.6\, {\rm pc}^{-1} \mbox{ at } \theta = \upi /6 \mbox{ and }15.6< k< 41.8\, {\rm pc}^{-1} \mbox{ at } \theta = \upi /3)$. Beyond the `gap' at large wave numbers, the slow magnetosonic wave speed appreciably increases. FMSW (curves 3, 4) remains unaffected by the transition into the thermally unstable region of parameter space and
behave the similar way as in  the cases depicted in Figs.~\ref{fig:7030}--\ref{fig:7060}.

\begin{table}
\caption{\label{tab:k}Critical wave numbers $k_{{\rm cr}}$ (in units of pc$^{-1}$) of the slow magnetosonic waves for
various values of $T$ and $\theta$.}
\begin{center}
\begin{tabular}{ c|c|c|c|c|c|c|c|c }
\hline
&10$^\circ$&20$^\circ$&30$^\circ$&40$^\circ$&50$^\circ$&60$^\circ$&70$^\circ$&80$^\circ$ \\
\hline
30 K& 14.7  & 16.3 & 18.6 & 22.0 & 27.1 & 35.8 & 53.4 & 106.3 \\
40 K& 12.9 & 14.1 & 16.1 & 19.0 & 23.4 & 30.8 & 45.9 & 91.4 \\
50 K& 11.2 & 12.0 & 13.4 & 15.7 & 19.2 & 25.2 & 37.3 & 74.3 \\
60 K& 10.2 & 10.9 & 12.1 & 13.9 & 16.9 & 22.1 & 32.7 & 64.9 \\
70 K& 10.3 & 10.8 & 11.9 & 13.6 & 16.5 & 21.4 & 31.5 & 62.5 \\
80 K& 12.4 & 13.0 & 14.2 & 16.1 & 19.4 & 25.0 & 36.7 & 72.6 \\
\hline
\end{tabular}
\end{center}
\end{table}

For the small wave numbers SMSW have faster growth rate than FMSW for $\theta = \upi /6$ (Figs. \ref{fig:7030}, \ref{fig:9530}) and
vice versa for $\theta = \upi /3$ (Figs. \ref{fig:7060}, \ref{fig:9560}) which proves the asymptotic analysis for the $A \approx 1$ or $\beta \approx 2$ (see Fig. \ref{fig:Q}).

The dependencies of the growth rates of the unstable
slow magnetosonic  and isobaric modes on the wave number are shown in Fig.~\ref{fig:slowdyn}(a)
for various values of temperature in
the range 35 - 120~K. Corresponding
frequencies are plotted in Fig.~\ref{fig:slowdyn}(b). The dependences are calculated for $\theta = \upi /6$. Fig.~\ref{fig:slowdyn}(a) shows that the maximum growth rates
of the unstable oscillatory and dynamical modes
increase with temperature. At $T=35$~K the minimal growth time of the slow
magnetosonic mode is $\sim 5\cdot 10^8$~yr at the wavelength $\lambda\sim 0.1$~pc.
At $T=70$~K the corresponding values are $\sim 2\cdot 10^7$~yr and $\sim 0.15$~pc, respectively.

In the range of temperatures 35-70~K, the slow magnetosonic
instability develops if the wave number is less than some threshold
value $k_{{\rm cr}}$ in accordance with Equation (\ref{kcrfs2}). Critical wave number $k_{{\rm cr}}\approx 5-10$~pc$^{-1}$ in the cases depicted in Fig.~\ref{fig:slowdyn}. The growth rate
increases with increasing temperature.
At $T=92$~K, the instability
has a character of standing waves with growth time $\approx 1.1\cdot 10^6$~yr at the small wavelengths. At $T=95$~K (solid line with rounds in Fig.~\ref{fig:slowdyn}), two slow magnetosonic waves transform to the dynamical  modes with different growth rates for $k\ga 10$~pc$^{-1}$.

Our calculations show that the branching point, where two dynamical modes transform into the
pair of the oscillatory
ones, shifts towards the smaller wave numbers, as
the temperature increases and reaches zero at $T\approx100$~K.
At $T=120$~K, the unstable slow magnetosonic modes disappear completely, and
there remains only one unstable isobaric mode (solid line with triangles in Fig.~\ref{fig:slowdyn}).

In Fig.~\ref{fig:fast}, we plot the dependence of the growth rates of FMSW on $k$ for various $T$ in the unstable region. The curves are plotted for $\theta = \upi /6$.
The growth rates of FMSW are smaller in comparison with the ones of SMSW. FMSW achieve their
maximum growth rates at $\lambda \approx 0.2-1$~pc with corresponding
growth time approximately $5\cdot 10^8$~yr.
The unstable FMSW do not transform into
isobaric instability in comparison with the SMSW modes.

In Fig.~\ref{fig:slowmax}, we plot the maximum growth rates of the unstable slow magnetosonic modes (panel a) and corresponding
wave numbers (panel b) versus temperature for
various values of the angle $\theta$. Fig.~\ref{fig:slowmax} shows that the maximum growth rate of
the SMSW modes very weakly depends on $\theta$ and grows with
growing temperature. The instability of SMSW exists only at $T>25$~K. At $T>92$~K, $k_{{\rm max}} \rightarrow\infty$ that corresponds to the transformation of the unstable SMSW into the
dynamical ionization-thermal modes.
The wave number $k_{{\rm max}}$ increases with
$\theta$, as Fig.~\ref{fig:slowmax}(b) shows.

\begin{table*}
\caption{\label{tab:sigma}Dependence of maximum increments, minimum growth times and corresponding wavelengths of acoustic waves and SMSW on the characteristics of the medium.}
\begin{center}
\begin{tabular}{c|c|c|c|c|c|c|c}
\hline
B, $\umu$G&$\xi_{{\rm CR}}$, s$^{-1}$ &$T$, K & $n$, cm$^{-3}$ & $x$&$\sigma _{{\rm max}}$, yr$^{-1}$ &$\sigma^{-1}_{{\rm max}}$, Myr  & $\lambda_{{\rm max}}$, pc\\
(1) & (2) & (3) & (4) & (5) & (6) & (7) & (8)\\
\hline
\multirow{4}{*}{0}& \multirow{2}{*}{$1\cdot 10^{-17}$}  & 50 & 42 & $2\cdot 10^{-4}$ & $2.1\cdot 10^{-8}$&$48$ & 0.94 \\
                  &                     & 70 & 24 & $3\cdot 10^{-4}$ &$6.0\cdot 10^{-8}$&$17$ & 1.05  \\
                  & \multirow{2}{*}{$5\cdot 10^{-17}$}  & 50 & 36 &$5\cdot 10^{-4}$ &$8.9\cdot 10^{-8}$& $11$ & 0.45 \\
                  &                     & 70 & 20 & $7\cdot 10^{-4}$ &$2.4\cdot 10^{-7}$&$4.1$ & 0.50 \\

\multirow{4}{*}{2}& \multirow{2}{*}{$1\cdot 10^{-17}$}  & 50 & 42 & $2\cdot 10^{-4}$ &$1.6\cdot 10^{-8}$& $63$ & 0.57 \\
                  &                     & 70 & 24 & $3\cdot 10^{-4}$ &$5.6\cdot 10^{-8}$&$18$ & 0.69  \\
                  & \multirow{2}{*}{$5\cdot 10^{-17}$}  & 50 & 36 &$5\cdot 10^{-4}$ &$7.1\cdot 10^{-8}$& $14$ & 0.25 \\
                  &                     & 70 & 20 & $7\cdot 10^{-4}$ &$2.2\cdot 10^{-7}$&$4.5$ & 0.31 \\

\multirow{4}{*}{6}& \multirow{2}{*}{$1\cdot 10^{-17}$}  & 50 & 42 &$2\cdot 10^{-4}$ &$2.0\cdot 10^{-8}$& $50$ & 0.63 \\
                  &                     & 70 & 24 &$3\cdot 10^{-4}$ &$5.9\cdot 10^{-8}$& $17$ & 0.75 \\

                  & \multirow{2}{*}{$5\cdot 10^{-17} $}  & 50 & 36 &$5\cdot 10^{-4}$ &$8.3\cdot 10^{-8}$& $12$ & 0.31  \\
                  &                     & 70 & 20 &$7\cdot 10^{-4}$ &$2.4\cdot 10^{-7}$& $4.2$ & 0.38 \\
\hline
\end{tabular}
\end{center}
\end{table*}

In Table \ref{tab:sigma}, we show maximum growth rate, $\sigma _{{\rm max}}$ (column 6), minimum growth time, $\sigma^{-1}_{{\rm max}}$ (column 7), and corresponding characteristic wavelengths $\lambda_{{\rm max}}$ of acoustic waves ($B=0$) and SMSW (column 8) for various magnetic field strengths (column 1) and cosmic ray ionization rates (column 2). Corresponding equilibrium values of density and ionization fraction are listed in columns 3 and 4, respectively.
Values presented in Table \ref{tab:sigma} are calculated for $\theta=\pi/4$.
Table \ref{tab:sigma} shows that magnetic field decreases the growth rates of SMSW as compared to the unstable acoustic waves.
The growth time of SMSW decreases with increasing magnetic field strength and cosmic ray ionization rate.
The growth time is of $4\cdot 10^6$~yr for $\xi_{{\rm CR}}=5\cdot 10^{-17}$ and $T=70$ K.

\begin{figure}
	\includegraphics[width=\columnwidth]{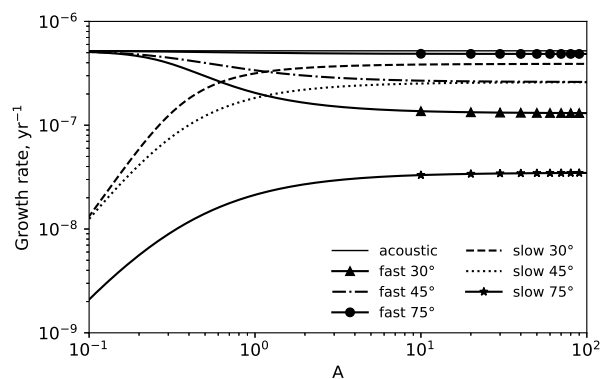}
   \caption{The dependences of the increments of FMSW and SMSW on the Alfv\`en number for various angles $\theta$, $T=70$~K, $h= 5\cdot 10^{-26}$~erg~s$^{-1}$, $\xi _\textrm{CR} = 5\cdot 10^{-17}$~s$^{-1}$, $k=10$~pc$^{-1}$. Horizontal solid line depicts the increment of the unstable acoustic wave.}
    \label{fig:incr}
\end{figure}

Dependence of $\sigma _{{\rm max}}$ on magnetic field strength can be analysed using asymptotic formulae from Section~\ref{Sec:ksmall}. The growth rates of SMSW and FMSW are the real parts of $\sigma _{{\rm s} \pm}$ and $\sigma _{{\rm f} \pm}$ in Equations (\ref{asympslowlinf}-\ref{asympfastlinf}). In Figure~\ref{fig:incr}, we plot dependences of these growth rates on the Alfv\`en number for various angles $\theta$, $T=70$~K, $h= 5\cdot 10^{-26}$~erg~s$^{-1}$, $\xi _\textrm{CR} = 5\cdot 10^{-17}$~s$^{-1}$, $k=10$~pc$^{-1}$. The growth rate of the unstable acoustic wave is also shown, which corresponds to the case $B=0$. The growth rate of the acoustic wave does not depend on $A$ and equals $5\times 10^{-7}$~yr$^{-1}$. Figure~\ref{fig:incr} shows that the growth rates of SMSW and FMSW are less than the growth rate of the acoustic wave, which reflects stabilizing effect of the magnetic field. The growth rates of SMSW and FMSW have similar behaviour as factors $Q_{{\rm s}}$ and $Q_{{\rm f}}$ depicted in Figure~\ref{fig:Q}. The growth rate of SMSW rapidly increases with $A$ for the case of a weak magnetic field ($A\ll 1$) and asymptotically tends to constant value for a strong magnetic field ($A\gg 1$). The growth rate of FMSW decreases with $A$ at $A\ll 1$ and is nearly constant at $A\gg 1$. Increase of the growth rate of SMSW with magnetic field strength is explained by the fact that the perturbation of magnetic  pressure has opposite sign as compared to the perturbation of gas pressure in SMSW~\citep[see, for example,][]{somov_book}. Therefore, the magnetic pressure of SMSW does not prevent pressure and density increase at the compression phase. On the contrary, magnetic pressure and gas pressure add up in FMSW, and the magnetic field prevents plasma compression. The growth rate of SMSW decreases with increasing angle $\theta$. SMSW do not propagate in the direction perpendicular to the magnetic field. For small angles $\theta$ and $A>1$, the growth rate of SMSW tends to the growth rate of the acoustic wave.  The growth rate of FMSW increases with the angle $\theta$.

\section{Conclusion}

We investigated the stability of the CNM with frozen-in magnetic field with the help of small perturbations. It was considered that the gas is heated by cosmic rays,  X-rays and electronic photoemission from the dust grains, while cooling is provided by collisional excitation of fine structure levels C~\textsc{ii} with hydrogen atoms and electrons. Ionization state of the CNM was determined from the balance between cosmic ray ionization and radiative recombinations.

Derived dispersion relation describes all modes of thermal instability \citep{field1}, and in particular thermal-reactive
modes \citep{yoneyama,corbelli}, ionization-coupled acoustic modes \citep{flannery}. We focused on the instability of magnetosonic waves, that arise in presence of magnetic field, in thermally stable region, with temperature $35<T<95$~K and density $n\la 10^3\,\mbox{cm}^{-3}$. The instability is affected by heating/cooling and ionization/recombination processes, so we called it as a magnetic ionization-thermal instability (MITI).

The dispersion equation is investigated
 analytically in the cases of small and large wave numbers,
as well as numerically in general case. Typical Galactic magnetic field strength,
ionization rates by cosmic rays and the rates of heating by
cosmic rays, X-rays and electronic photoemission from the dust grains are adopted in the calculations.

The asymptotic analysis and numerical solution of the dispersion equation show that MITI has the threshold behaviour in the CNM. The magnetosonic modes can be
unstable only at the wave number less than a critical one, $k < k_{{\rm cr}}$. We obtain the expressions for the
critical wave numbers, $k_{{\rm cr}}$, of the slow and fast magnetosonic modes. The critical wave number for the slow magnetosonic waves increases with the angle $\theta$ between the magnetic field and wave vector. Typical  values of critical wave number $k_{{\rm cr}}$ lie in range $2-20$~pc$^{-1}$ depending on $\theta$ and gas temperature.

In the limit of large wavelengths, the unstable slow magnetosonic waves have larger growth rate than the fast magnetosonic waves for the angle between magnetic field lines
and wave vector $\theta<\upi/4$ and Alfv\`en number $A\gg1$.
In the isobarically unstable region ($T \geq 100$ K) the slow magnetosonic
modes are stable, but the unstable fast
magnetosonic modes have small growth rates $\sigma_{\rm R} \la 10^{-9}$ yr$^{-1}$ and do not play significant role
in the dynamics of medium.

Depending on the angle $\theta$ in the range from
0 to $\upi /2$, the slow magnetosonic
modes have the most unstable wavelengths $\lambda _{{\rm max}}\approx 0.1-0.15$~pc.
The growth time of the unstable slow magnetosonic
modes decreases with magnetic field strength and cosmic ray ionization rate. It lies in range from $\sim 4$ to $\sim 60$~Myr. For example, the growth time is of $\sim 4$~Myr for $B=6\,\mu$G, $\xi_{{\rm CR}}=5\cdot 10^{-17}\,{\rm s}^{-1}$ and $T=70$~K. This time is less than the characteristic time of Galaxy spiral pattern rotation period of about $\approx 200$~Myr in the solar neighbourhood~\citep[according to data from][]{reid14}. Therefore, MITI can develop over the dynamical time of Galaxy evolution.
The formation of clouds in the CNM is often explained by the action of the interstellar turbulence and shock waves~\cite[see review][]{elmegreen04}.
The turbulent velocity at the scale $l_0$ can be estimated as $v_{{\rm turb}}=1\,(l_0/$1 pc$)^{\delta}$~km~s$^{-1}$, where $\delta \sim 0.4-0.6$~\citep[see e.g.][]{larson81,dudorov91,maclow,ballesteros,kritsuk13}.
Therefore, the characteristic time-scale of the interstellar turbulence $\tau_0=l_0/v_{{\rm turb}}$  is of order of few Myr in diffuse clouds with sizes $l_0 < 10$~pc
and  $\tau_0 \ga 10^7$ yr for clouds with $l_0 \ga 10$~pc.
These times are comparable to the growth time of MITI.
We propose that MITI can lead to the formation of condensations, such as TSAS, in the diffuse CNM, along with the turbulence. The condensations would have sizes more than $0.05-0.5$~pc under typical conditions in the CNM.

The growth of the slow magnetosonic waves with the wave vector almost perpendicular
to the magnetic field lines, $\theta =\upi/2$, can lead to the formation of condensations elongated in
the direction of magnetic field. This instability can be one of  the possible mechanisms of the formation of filament-like clouds with typical width of 0.1 pc threaded by parallel magnetic field in the ISM~\citep[see][]{andre14, planck16_II}.

We also propose that MITI at the non-linear stage can lead to the generation of MHD wave turbulence. The dependence of velocity dispersion on spatial scale has the from of power law
$\Delta v \propto l^{1/2}$ in this case~\citep{sagdeev}. That type of turbulence can one of the possible mechanisms of the condensations of TSAS. This suggestion must to be checked with 3-D numerical modeling of non-linear stage of MITI.
We plan to study the development of MITI at higher densities and stronger magnetic field in future, to investigate the role of MITI in interstellar molecular clouds and accretion disks of young stars. We also suggest that mechanism of MITI can be applied to H \textsc{ii} regions.

\section*{Acknowledgements}

The authors are thankful to the referee prof. Robi Banerjee for useful comments that helped us to improve the quality of this work.
The work of A.E. Dudorov and S.O. Fomin is supported partially by development program of the Chelyabinsk State University.
The work of S.~Khaibrakhmanov in Section~3 is supported by the Russian Foundation for Basic Research (\textnumero1802-01067/18), and the work in Section 4 is supported by the Russian Science Foundation (project 18-12-00193).




\bibliographystyle{mnras}
\bibliography{miti} 

\end{document}